\documentclass{article}

\usepackage{arxiv}
\usepackage[utf8]{inputenc} 
\usepackage[T1]{fontenc}    
\usepackage{hyperref}       
\usepackage{url}            
\usepackage{booktabs}       
\usepackage{amsfonts}       
\usepackage{nicefrac}       
\usepackage{microtype}      
\usepackage{lipsum}		
\usepackage{graphicx}
\usepackage{natbib}
\usepackage{doi}
\usepackage{comment}
\usepackage[nocompress]{cite}

\usepackage{times}
\usepackage{booktabs} 
\usepackage{paralist}
\usepackage{multirow}
\newcommand{\cut}[1]{}
\usepackage{graphicx}
\usepackage{color}
\usepackage{amsmath}
\usepackage{amssymb}

\usepackage{url}
\usepackage{wrapfig}
\usepackage{algpseudocode}
\usepackage{algorithm}
\usepackage{algcompatible}
\usepackage{epstopdf}
\usepackage{multirow}
\usepackage{rotating}
\usepackage{relsize}
\usepackage{appendix}
\usepackage{bbm}
\usepackage{mathtools}

\title{Querying Triadic Concepts through Partial or Complete Matching of Triples}

\author{ \href{https://orcid.org/0000-0001-6423-8681}{\includegraphics[scale=0.06]{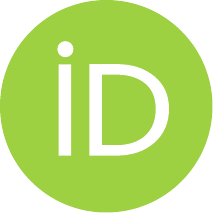}\hspace{1mm}Pedro Henrique B. Ruas} \\
	Department of Computer Science and Engineering \\ 
         University of Quebec in Outaouais \\
         Gatineau, Quebec, Canada\\
	\texttt{pedrohbruas@gmail.com} \\
	\And
	\href{https://orcid.org/0000-0001-7410-4177}{\includegraphics[scale=0.06]{orcid.pdf}\hspace{1mm}Rokia Missaoui}\thanks{ http://w3.uqo.ca/missaoui/} \\
	Department of Computer Science and Engineering \\ 
         University of Quebec in Outaouais \\
         Gatineau, Quebec, Canada\\
	\texttt{rokia.missaoui@uqo.ca} \\
       \And
	\href{https://orcid.org/0000-0002-0604-2709}{\includegraphics[scale=0.06]{orcid.pdf}\hspace{1mm}Mohamed Hamza Ibrahim} \\
	Department of Computer Science and Engineering \\ 
         University of Quebec in Outaouais \\
         Gatineau, Quebec, Canada\\
	\texttt{mohamed.ibrahim@polymtl.ca} \\
}

\renewcommand{\shorttitle}{Querying Triadic Concepts through Partial or Complete Matching of Triples}
\def\neu#1{{\em #1}}
\def\KK{\mathbb{K}}

\begin{document}
\maketitle

\begin{abstract}
In this paper, we introduce a new method for querying triadic concepts through partial or complete matching of triples using an inverted index, to retrieve already computed triadic concepts that contain a set of terms in their extent, intent and/or modus. 
As opposed to the approximation approach described in Ananias, this method (i) does not need to keep the initial triadic context or its three dyadic counterparts, (ii) avoids the application of derivation operators on the triple components through context exploration, and (iii) eliminates the requirement for a factorization phase to get triadic concepts as the answer to one-dimensional queries. 
Additionally, our solution introduces a novel metric for ranking the retrieved triadic concepts based on their similarity to a given query. Lastly, an empirical study is primarily done to illustrate the effectiveness and scalability of our approach against the approximation one. Our solution not only showcases superior efficiency, but also highlights a better scalability, making it suitable for big data scenarios.
\end{abstract}

\keywords{Data mining \and Triadic Concept Analysis \and Querying Triadic Concepts
\and Triple Matching}

\section{Introduction}
\label{Intro}

In information systems, users can be overwhelmed by data and even patterns (knowledge), yet they are often interested by specific knowledge nuggets or would like to find some specific ones. The scope for exploring patterns can vary widely among users and often evolves over time, with a common preference for an exploratory and iterative process that uncovers patterns using relational operations on lattices such as selection and projection \citep{Kwuida2010}.

In this particular context, Formal Concept Analysis (FCA) presents a valuable mathematical framework for representing and extracting knowledge from data. The inception of FCA dates back to the 1980s when Wille and Ganter first proposed it \citep{Ganter1999} as a framework for constructing and exploring concept lattices and extracting association rules. 

Despite its success in various applications, the classical approach may not always be sufficient, and some situations require an extension with an additional dimension to obtain a more complete characterization and representation of the data. \citep{Lehmann1995} took the initiative to extend FCA in their work, introducing Triadic Concept Analysis (TCA) to describe a ternary relationship among object, attribute, and condition sets. An instance where such an extension is beneficial is in a social resource sharing system, commonly referred to as a \textit{folksonomy} \citep{Jaschke2008}, where users, resources, and keywords (tags) establish a ternary relationship through users' annotations. In their paper, \citep{Lehmann1995} proposed a three-dimensional graphical representation known as a trilattice.

A few researchers have attempted to simplify the visualization and navigation through triadic concepts \citep{Missaoui2020,Rudolph2015} by proposing navigation strategies and graphical representations based on the classical Hasse diagram.

In our previous work \citep{Missaoui2020}, we introduced the \textit{T-iPred} algorithm, which is an adaptation of the \textit{iPred} algorithm \citep{Baixeries2009} used for efficiently computing links between concepts in FCA. This adaptation aims to represent the Hasse diagram of triadic concepts, and its graphical representation facilitates the exploration and discovery of hidden knowledge within triadic contexts.

However, when exploring triadic concepts through the Hasse diagram, it is common to encounter a diagram with hundreds or even thousands of triadic concepts. Indeed, yet triadic contexts with a reduced number of objects, attributes, and conditions can produce a vast amount of triadic concepts.

In this regard, manually exploring triadic concepts can become an impractical task due to the large number of concepts involved. It would require the user to navigate through the entire diagram to analyze the concepts of interest. In this study, we propose an approach based on an inverted index which allows users to make queries on all triadic concepts. Users can search for a triple $(A_1, A_2, A_3)$, and if the query corresponds to a triadic concept, the platform will return this concept. In case the searched triple is not a triadic concept, a set of more similar triadic concepts will be returned. Furthermore, our solution enables users to conduct searches through queries specifying only one dimension (objects \textit{or} attributes \textit{or} conditions), two dimensions (objects \textit{and} attributes, objects \textit{and} conditions, or attributes \textit{and} conditions), or all three dimensions.

The structure of this article is as follows: Section \ref{Background} provides an overview of the theoretical foundation of Triadic Concept Analysis and inverted index, whereas Section \ref{Related_work} presents the related work. Section \ref{Triple_search} presents the proposed approach to query triadic concepts. Section \ref{Results} describes an empirical study using both synthetics and real data sets. A conclusion and further investigations are given in Section \ref{Conclusion}.

\section{Background}
\label{Background}

Within this section, we aim at introducing fundamental definitions of Formal Concept Analysis and Triadic Concept Analysis. Our primary focus will be on the triadic approach.

\subsection{Formal Concept Analysis}

Formal Concept Analysis (FCA) was introduced in \citep{Ganter1999} as a branch of applied mathematics, which is based on a formalization of concept and concept 
hierarchy. It starts from a formal binary context $\KK:=(\mathcal{G}, \mathcal{M}, \mathcal{I})$ where $\mathcal{G}$, $\mathcal{M}$ and $\mathcal{I}$ are a set of objects, a set of attributes, and a binary relation between $\mathcal{G}$ and
$\mathcal{M}$ respectively , to construct a concept (Galois) lattice whose nodes are formal concepts (maximal rectangles) described by an extent (set of objects) and an intent (set of attributes).

Given arbitrary subsets $A \subseteq \mathcal{G}$ and $B \subseteq \mathcal{M}$, the following derivation operators are defined:\\
$A^{\prime{}} = \{m \in \mathcal{M} \mid \forall g \in A, (g,m) \in \mathcal{I} \}, \; A \subseteq \mathcal{G}$ and \\ 
   $B^{\prime{}} = \{g \in \mathcal{G} \mid \forall m \in B, (g,m) \in \mathcal{I} \}, \; B \subseteq \mathcal{M} $ \\
where $A^{\prime{}}$ is the set of attributes common to all objects of $A$ and $B^{\prime{}}$ is the set of objects sharing all attributes from $B$. 

The pair $c=(A,B)$ is called a \textit{formal concept} of $\mathbb{K}$ with \textit{extent} $A$ and \textit{intent} $B$ if $A^{\prime{}}=B$, and $B^{\prime{}}=A$.

A partial order $\preceq$ exists between two concepts $c_1=(A_1,B_1)$ and $c_2=(A_2,B_2)$ $\iff A_1 \subseteq A_2$ (or, equivalently, $B_2 \subseteq B_1$). The set $\mathcal{C}$ of all concepts together with the partial order form a concept lattice.

\subsection{Triadic Concept Analysis}


Triadic Concept Analysis (TCA) was initially introduced by Lehmann and Wille \citep{Lehmann1995, Wille95} as an extension of Formal Concept Analysis \citep{Ganter1999}. It serves to analyze data described by three sets: $K_1$ (objects), $K_2$ (attributes), and $K_3$ (conditions), along with a ternary relation $Y\subseteq K_1{\times}K_2{\times}K_3$. This structure is named a \neu{triadic context} and denoted as $\KK:=(K_1, K_2, K_3, Y)$. An example of such a context is illustrated in Table \ref{tab:PNRKST_context}, representing customers' purchases in $K_1=\{1,2,3,4,5,6\}$ from suppliers in $K_2= \{\textbf{P}eter, \textbf{N}elson, \textbf{R}ick, \textbf{K}evin, \textbf{S}imon, \textbf{T}revor\}$ of products in $K_3 =\{\textbf{a}ccessories, \textbf{b}ooks, \textbf{c}omputers, \textbf{d}igital \ cameras\}$.

\begin{table}[!ht]
	\begin{center}
	\begin{footnotesize}
	    \begin{tabular}{|c|c|c|c|c|c|c|}
	      \hline
	      & $P$ & $N$ & $R$ & $K$ & $S$ & $T$ \\
	      \hline
	      $1$ &  abd  &  abd  &  ac &  ab &  a &  a  \\
	      $2$ &  ad  &  abcd  &  abd &  ad &  ad &  a \\
	      $3$ &  abd  &  ad  &  ab &  ab &  a   &  a\\
	      $4$ &  abd  &  abd  &  ab &  ab &  ad  &  a\\
	      $5$ &  ad  &  ad  &  abd &  abc &  a &  ab \\
            $6$ & abcd & abcd & abcd & abcd & abcd & abcd \\
	      \hline
	    \end{tabular}  
        \caption{A triadic context}
		 \label{tab:PNRKST_context}
	\end{footnotesize}
	\end{center}
\end{table}


The notation $(a_1, a_2, a_3) \in Y$ indicates that the object $a_1$ possesses the attribute $a_2$ under the condition $a_3$. For instance, the value $ac$ at the intersection of Row 1 and Column $R$ means that Customer 1 orders products $a$ and $c$ from Supplier $R$.

The triadic context $\KK:= (K_{1}, K_{2}, K_{3}, \textit{Y})$ can be converted into three dyadic contexts as follows.

\begin{align*}
    \begin{split}
        \KK^{(1)} :=  (K_{1}, K_{2} \times K_{3}, Y^{(1)}) \ \textrm{with} \ a_{1}Y^{(1)}(a_{2}, a_{3}) \Leftrightarrow (a_{1}, a_{2}, a_{3}) \in Y\\
        \KK^{(2)} :=  (K_{2}, K_{1} \times K_{3}, Y^{(2)}) \ \textrm{with} \ a_{2}Y^{(2)}(a_{1}, a_{3}) \Leftrightarrow (a_{1}, a_{2}, a_{3}) \in Y\\
        \KK^{(3)} :=  (K_{3}, K_{1} \times K_{2}, Y^{(3)}) \ \textrm{with} \ a_{3}Y^{(3)}(a_{1}, a_{2}) \Leftrightarrow (a_{1}, a_{2}, a_{3}) \in Y\\
    \end{split}
\end{align*}



A \emph{Triadic Concept} (TC) or \emph{closed tri-set} within a triadic context $\KK$ is a triple $(A_1, A_2, A_3)$ with $A_1 \subseteq K_1$, $A_2 \subseteq K_2$, $A_3 \subseteq K_3$, and $A_1{\times}A_2{\times}A_3\subseteq Y$ that is \textit{maximal} in $Y$. In other words, none of the three subsets can be expanded without violating the ternary relation $Y$. Thus, this triple represents a maximal cuboid filled with ones (or crosses). For instance, the tri-set $(5\,6, K\,R, a\,b)$ is not closed because $5\,6\times K\,R\times a\,b \subsetneq 3\,4\,5\,6\times K\,R \times a\,b \subseteq Y$. We use $(A_1, A_2, A_3)$ or $A_1\times A_2\times A_3$ interchangeably to refer to a triadic concept.

The terms \emph{extent}, \emph{intent} and \emph{modus} refer to $A_1$, $A_2$ and $A_3$ of the concept, respectively. We propose to name the pair $({A}_{2},{A}_{3})$ as the \textit{feature} associated with $A_1$.


To compute triadic concepts, two derivation operators are introduced. Let $\KK:=(K_1, K_2, K_3, Y)$ be a triadic context, and let $\{i, j, k\} = \{1, 2, 3\}$ with $j < k$. Consider $X_i \subseteq K_i$ and $(X_j, X_k) \subseteq K_j {\times} K_k$\footnote{We use $(X_j, X_k) \subseteq K_j {\times} K_k$ to indicate that $X_j\subseteq K_j$ and $X_k \subseteq K_k$.}. The {$^{(i)}$-derivation} \citep{Lehmann1995} is defined as follows:

\vspace{-.5cm}
\begin{align*}
    X_i^{(i)} &:= \{(a_j, a_k) \in K_j {\times} K_k \mid (a_i, a_j, a_k) \in Y ~\forall a_i \in X_i\}\\
   (X_j, X_k)^{(i)} &:= \{a_i \in K_i \mid (a_i, a_j, a_k) \in Y~{\rm for~all}~(a_j, a_k) \in X_j{\times}X_k\}
\end{align*}


As an illustration, the application of the first derivation operator on the set $3456$\footnote{Throughout this paper, we will frequently employ simplified notations for sets. For instance, $1\,2\,5$ (or simply $125$) represents the set $\{1, 2, 5\}$, while $a\,b$ (or simply $ab$) denotes $\{a,b\}$.} leads to two concept features $(K, ab)$ and $(R, ab)$ after the factorization of the pairs $(K, a)$, $(K, b)$, $(R, a)$ and $(R, b)$,  i.e., after the computation of maximal rectangles involving attributes and conditions in this example. Similarly, by applying the second derivation operator on the pair $(KPR, ab)$, we find the extent $346$, meaning that only Customers 3, 4 and 6 buy Products $a$ and $b$ from Suppliers $K$, $P$ and $R$.



Once computed, the set of triadic concepts grouped under the same extent can be ordered using the order relations $\leq_{1}$ determined by the quasi-order $\lesssim_{1}$ to form a poset. This sorting creates a Hasse diagram where each node represents all the triadic concepts with the same extent. However, it is important to note that the generated nodes do not constitute a complete lattice since the intersection of extents is not necessarily an extent in the triadic setting.

We recall that for two elements $x$ and $y$ in a poset, if $x\le y$ (resp. $x < y$), then $x$ is below (resp. strictly below) $y$. If $x<y$ and there is no element between $x$ and $y$, $x$ is called a lower cover of $y$, and $y$ an upper cover of $x$, and we write $x\prec y$.

\begin{figure*}[ht]
\centering
\includegraphics[scale=.4]{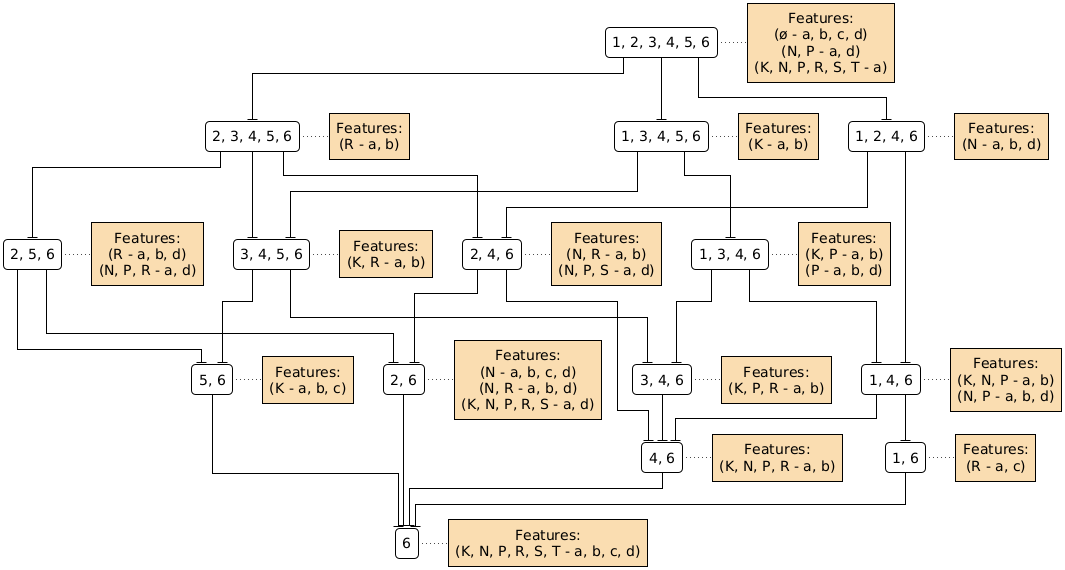}
\caption{The Hasse diagram of triadic concepts}
\label{fig:hasse_diagram}
\end{figure*}

Figure \ref{fig:hasse_diagram} presents the Hasse diagram generated by the \textit{T-iPred} algorithm, where the value inside each node represents an extent while the pairs of values attached to a dotted line are the corresponding features. For instance, the node labelled $2 5 6$ encompasses the extent $2 5 6$ of the TCs $(256, R, abd)$ and $(256, NPR, ad)$.

\subsection{Inverted index}

An inverted index is a powerful data structure commonly used in information retrieval systems. Its primary purpose is to facilitate efficient and quick access to relevant information from large data collections. The main idea behind an inverted index is to map terms or keywords found in documents to the corresponding positions where these terms appear. Instead of looking for the entire document set for a specific term, the inverted index allows direct access to the documents containing that specific term, significantly reducing search time.

The benefits of utilizing an inverted index are manifold. Firstly, it greatly accelerates search operations, making it ideal for applications involving vast amounts of text data, such as search engines and document retrieval systems. Additionally, an inverted index supports diverse advanced search functionalities like Boolean queries, phrase searches, and ranking, enabling precise and context-aware retrieval \citep{Zobel2006}.


In the context of inverted index-based information retrieval, re-rank algorithms have a crucial role in fine-tuning the ranking of retrieved documents to provide users with more relevant results. After using the inverted index to quickly identify documents containing the query terms, the initial ranking may not fully capture the user's intent or context. Re-rank algorithms step in to reevaluate and adjust the document rankings based on additional features or criteria, improving the overall accuracy of the search results. By incorporating re-rank algorithms, information retrieval systems can enhance the user experience and deliver more accurate results.

\section{Related work}
\label{Related_work}

To the best of our knowledge, there is only one related work to our proposed approach. 
In a recent study, \citep{Ananias2021} introduced an approach to unveil patterns from triadic contexts by querying the computed triadic concepts. The proposed search strategy utilizes the diagram generated by \textit{T-iPred} to showcase the query's exact answer or the upper and lower covers for concepts that approximate the given query expressed as a triple of object, attribute, and condition sets. This methodology provides a powerful means to explore and extract valuable insights from triadic data structures.

In the approach proposed by \citep{Ananias2021}, the triadic context is first converted into three distinct dyadic contexts as indicated earlier. However, the computational cost and execution time to perform the Cartesian product among the two out of the three dimensions of the triadic context can be prohibitive even for a context with just a few dozen attributes and/or conditions. Furthermore, the derivation operation is performed several times during the process of querying triadic concepts, which can result in excessive processing time. Another limitation concerns the number of triadic concepts returned. By adopting derivation operators, the answer to queries often leads to one triadic concept in the scenario of an approximate search. This limitation prevents the user from exploring other triadic concepts that are, in fact, similar to the specified query and hence could be relevant to the user.

In \citep{Ananias2021}, when the user asks for a one-dimensional query $(X_{1}, -, -)$, $X_1'$ is first computed then factorized to get maximal rectangles as the set $\mathcal{F}$ of individual features $F_i = (X_{2i}, X_{3i})$ associated with $X_{1}$. Then, the set of the smallest triadic concepts that have at least $X_1 \subseteq K_1$ in their extent is calculated. For a two-dimensional query $(X_1, X_2, -)$, Proposition 3 in \citep{Wille95} is used to compute $(A_1, A_2, A_3)$ by first calculating $A_3$ and then $A_1$ followed by $A_2$ (or $A_2$ followed by $A_1$) using the second derivation on pairs (see Section \ref{Background}), leading to one or two triadic concepts. The case of three-dimensional queries $(X_{1}, X_{2}, X_{3} )$ is converted into three two-dimensional queries whenever the triple is not a triadic concept. 

Regarding the exploration of triadic concepts, the prototype known as ``FCA Tools Bundle" \citep{Kis2016} offers visualization and navigation capabilities through a set of triadic concepts by utilizing dyadic projections. It also aids in identifying concepts within a triadic context without computing all closed trisets. Nevertheless, the present paper investigates triple matching, which has not been examined in the mentioned prototype.
Another approach to approximating triadic concepts is through OAC-triclusters, akin to OA-biclusters approximating formal concepts by relaxing the closed triset properties. However, no specific query formulation is pursued in the quest for triadic concepts that precisely or partially match a given input triple.

\section{Partial or Complete Matching of Triples}
\label{Triple_search}

In this section, we present a new approach for extracting knowledge from triadic contexts through partial or complete matching of triples. The process of querying triples within the TCA framework involves searching for patterns in the form of triadic concepts, utilizing our proposed approach that requires only the triadic concepts as input.

Our primary objective with this solution is to identify all the triadic concepts that either match the user's query completely or partially. For instance, consider the triadic context of customers represented in Table \ref{tab:PNRKST_context}. 
This context captures transactions between customers and suppliers over products. A specialized professional might be interested in determining if there exists a set of products bought by a specific group of customers from suppliers. For example, the formal concept $(346, KPR, ab)$ denotes that customers 3, 4, and 6 purchased the same products \textit{a} and \textit{b} from the suppliers $K$, $P$, and $R$.



If one seeks concepts with an extent equal to $\{1 4 6\}$, the resulting set would be $(146, KNP, ab)$ and  $(146, NP, abd)$. 
Conversely, specifying the set $\{1, 3\}$ reveals that there is no concept with just these two objects in the Hasse diagram (see Figure \ref{fig:hasse_diagram}). This indicates that Customers 1 and 3 likely made purchases alongside other ones.
However, there are six triadic concepts in which the set $\{1, 3\}$ is a subset of the extent : $(1346, KP, ab)$, $(1346, P, abd)$, $(13456, K, ab)$, $(123456, NP, ad)$, $(123456, KNPRST, a)$ and $(123456, \varnothing, abcd)$. These concepts provide valuable information about the purchases that the two clients share with other costumers. 



Another user might be interested in triadic concepts that partially match the elements in the query, in case there is no exact match. For instance, when searching for $\{1, 3\}$, we could return the concepts that have only one of the two elements in the query. In this scenario, the following triadic concepts would be returned to the user: $(16,R,ac)$, $(146,KNP,ab)$, $(146,NP,abd)$, $(346,KPR,ab)$, $(3456,KR,ab)$, $(1246,N,abd)$, $(23456,R,ab)$.


Both scenarios are possible to be executed in our approach. The proposed algorithm uses an inverted index that is created based on the triadic concepts, which enables not only an extremely efficient search, but also allows greater flexibility in the number of returned triadic concepts given a query. 

However, this flexibility can lead to a significant number of triadic concepts being returned, and in some cases, we might have several dozen concepts. In this regard, we also propose the \textit{Re-rank} algorithm to indicate the most similar triadic concepts to the user's query. Each returned concept is assigned a score based on the similarity of the query with the triadic concepts in the inverted index.

\begin{figure}[ht]
\centering
\includegraphics[scale=.35]{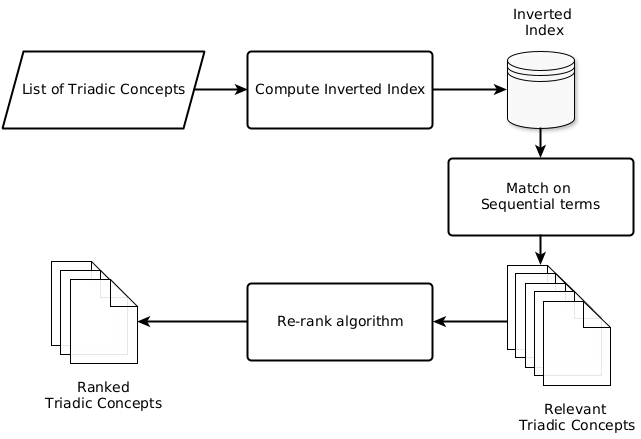}
\caption{Workflow of our proposed solution}
\label{fig:workflow}
\end{figure}


Figure \ref{fig:workflow} presents a diagram of how the proposed solution works. The initial step in the diagram involves comprehending all the triadic concepts. Subsequently, each constituent element of a triadic concept is scrutinized, and the inverted index is established, encompassing the mapping of elements to the concepts in which they appear.
When a user submits a query, the inverted index is queried for concepts possessing one or more elements present in the query. The assemblage of concepts that intersect with the query is referred to as “Relevant Triadic Concepts”.
Then, this set of triadic concepts is ranked based on their similarity with the query and then presented to the user.

\begin{figure*}[ht!]
\centering
\includegraphics[scale=.3]{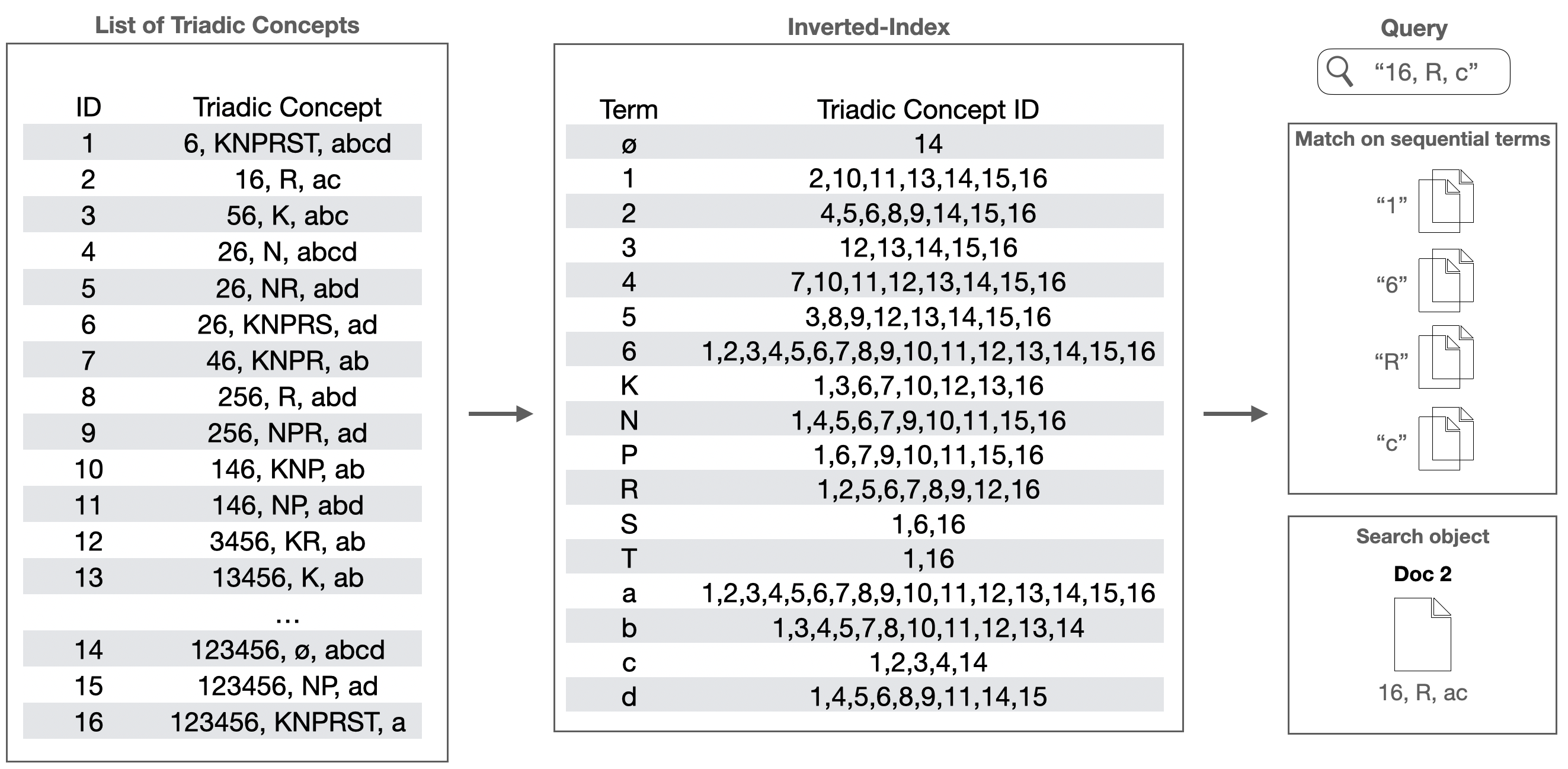}
\caption{Process of creating the Inverted-Index from triadic concepts and retrieving a matched concept}
\label{fig:inverted_index_tc}
\end{figure*}

For a better understanding of the proposed approach, we illustrate the first three steps of the workflow in Figure \ref{fig:inverted_index_tc}. On the left, we have the set of triadic concepts associated with a unique identifier. Then, in the center of the figure, each element (objects, attributes and conditions) composing the concepts is examined (terms), and an inverted index is created, indicating in which triadic concept (ID) each term occurs. This structure is used to perform matches based on a user's query. For example, when executing the query $(16, \ R, \ c)$ and submitting it to the inverted index, the response is the document with ID 2, corresponding to the concept $(16, \ R, \ ac)$.

The proposed metric is primarily based on the intuition that in the query ($X_1, X_2, X_3$), the component $X_i$ with the greatest number of elements should be assigned a greater weight than the other dimensions, indicating a likely higher level of interest.
For instance, when the user addresses the query $(-,\ R,\ a)$, the elements of the second and third dimensions of the query will carry equal weight in ranking the triadic concepts that include these elements. However, when entering the query $(-,\ R,\ abcd)$, the third dimension will hold a higher weight, as out of the five elements present in the query, four pertain to the conditions component (third dimension). Consequently, triadic concepts that exactly match the query's elements in the conditions ($abcd$) should possess a higher score than concepts that solely possess the attribute $R$ and/or a subset of the elements forming the set of conditions. The Re-rank procedure is presented in Algorithm \ref{alg:re-rank}.

\begin{tiny}
\begin{algorithm}[ht]
\caption{Re-rank algorithm}
\textbf{Input:}{ $Invert, Query, Tolerance \Theta$} \\
\textbf{Output:}{TC-Set: Ranked relevant TCs.}
\begin{algorithmic}[1] 
\STATE $TCs \gets getRelevantConcepts(Invert, Query, \Theta)$;
\STATE $X_1, X_2, X_3 \gets Query$;
\STATE $rankedConcepts \gets \left [  \right ]$\;

\FOR{$TC \ in \  TCs$}
\STATE $A_1, A_2, A_3 \gets TC$;  $\Delta A_1 \leftarrow \Delta A_2 \leftarrow \Delta A_3 \leftarrow 0$;
\STATE $\Delta A_1 \leftarrow 0$;
\STATE $\Delta A_2 \leftarrow 0$;
\STATE $\Delta A_3 \leftarrow 0$;

\IF{$|X_1| > 0$}
 \STATE $\Delta A_1 = |X_1 \cap A_1|$;
   \STATE     $\Delta A_1 \mathrel{+}= \frac{\Delta A_1} {max(|X_1|, |A_1|)}$;
 \ENDIF
\IF{$|X_2| > 0$}
 \STATE $\Delta A_2 = |X_2 \cap A_2|$;
   \STATE     $\Delta A_2 \mathrel{+}= \frac{\Delta A_2} {max(|X_2|, |A_2|)}$;
 \ENDIF

 \IF{$|X_3| > 0$}
 \STATE $\Delta A_3 = |X_3 \cap A_3|$;
   \STATE     $\Delta A_3 \mathrel{+}= \frac{\Delta A_3} {max(|X_3|, |A_3|)}$;
 \ENDIF

    \STATE $ranking = (\Delta A_1 \times \frac{|X_1|}{ |query|}) + (\Delta A_2 \times \frac{|X_2|}{ |query|}) +  (\Delta A_3 \times \frac{|X_3|}{ |query|})$\;
    \;
    \STATE $rankedConcepts.append(TC, ranking)$\;
\ENDFOR
\STATE \text{\textbf{Return}} \;TC-Set;
\end{algorithmic}
\label{alg:re-rank}
\end{algorithm}
\end{tiny}


In the first line of Algorithm \ref{alg:re-rank}, the procedure starts with three input: (i) the inverted index \textit{Invert} of terms (objects, attributes and conditions) and their corresponding concepts, (ii) the query and (iii) the tolerance parameter $\Theta$. 
The purpose of the  tolerance parameter is to allow the user to specify how many elements of the query might not be present in the returned triadic concepts. For instance, given the query $(156, -, -)$ and a tolerance parameter $= 0$, only the triadic concepts containing all the three objects $1$, $5$ and $6$ should be returned. If the tolerance parameter is $= 1$, triadic concepts that partially match the query with one missing element (here object) can also be returned, such as the concept $(5 6, K, abc)$.

In Line 2, the query is separated into three variables, each representing one of the three components of the triple. In Line 3, the variable that stores the triadic concepts and the calculated score is also initialized to an empty list. 
In Lines 4 to 23, all the triadic concepts returned after querying the inverted index undergo the Re-rank process. This process begins with the separation of the concept's components (Line 5). In Lines 6 to 8, three variables that will store the count of different elements between the user's query and the triadic concept are initialized to 0. 

In Line 9, it is checked whether the query includes elements from the first dimension (objects). If affirmative, in Line 10, the cardinality of the intersection between the query's objects and the elements that comprise the concept's extent is computed. Subsequently, in Line 11, this value is added to the result of dividing the variable $\Delta A_1$ by the maximum of $X_1$ and $A_1$.

The same reasoning is repeated for the second and third components of each triadic concept (Lines 13 to 20). In Line 21, the final score is calculated by summing up all the deltas and weighting them by the cardinality of each component in the query relative to the total length of the search string. To better illustrate the procedure, consider the query $(3456, -, a)$. Out of the five elements present in the query, four are from the first component (objects). Consequently, triadic concepts with a larger intersection with the query's elements in the first dimension will receive a higher score than concepts that only intersect in the third component (conditions).
Finally, in Line 23, the triadic concept along with the score are appended to the variable that stores the output.

\subsection{One-dimensional queries}

One-dimensional queries carry out the process of matching concepts by focusing on a single dimension from the three available. When dealing with a triple where only one dimension is given, the procedure of conducting a one-dimensional query entails the identification of triadic concepts that exhibit the highest similarity to the elements provided within the query.

Formally, a one-dimensional query can be precisely defined as a triple in which solely one dimension is known, and this approach can be applied to any of the three dimensions. In this context, three distinct triples are established: for the extent $(X_{1}, -, -)$, the intent $(-, X_{2}, -)$, or the modus $(-, -, X_{3})$.

\subsubsection{Example}

The list of concepts below contains the top 3 triadic concepts returned for the query $(-, KP, -)$.

\begin{enumerate}
    \item $(1346,\ KP,\ ab) - [3.0]$
    \item $(146,\ KNP,\ ab) - [2.67]$
    \item $(346,\ KPR,\ ab) - [2.67]$
\end{enumerate}

The first concept is the only one that includes all the elements specified in the query and, thus, has the highest value for the similarity metric ($3.0$). The second and third concepts have $KP$ in their intent. However, each one has an additional attribute ($N$ and $R$, respectively) and hence,  have the same similarity value.

For comparison, we executed the same query using the algorithm proposed by \citep{Ananias2021} and obtained the unique TC $(1346,\ KP,\ ab)$. We observe that both solutions found the same concept as the closest answer to the query. However, in our method, the user has the option to explore more similar concepts with their similarity score value.


\subsection{Two-dimensional queries}
A two-dimensional query facilitates either the identification of an existing concept or the detection of TCs that offer the closest match to that query. The objective is to uncover information linked to concepts that exhibit the highest similarity with any two components from the available three ones.

The formulation of a two-dimensional query involves a triple in which two dimensions are provided, while the third one remains unknown. Consequently, three distinct variants of this triple can be established: $(X_{1}, X_{2}, -) $, $(X_{1}, -, X_{3}) $, and $(-, X_{2}, X_{3})$.

\subsubsection{Example}

The top 3 concepts for the two-dimensional query $(-,\ R,\ ab)$ are as follows:

\begin{enumerate}
    \item $(23456,\ R,\ ab) - [2.67]$
    \item $(3456,\ KR,\ ab) - [2.5]$
    \item $(246,\ NR,\ ab) - [2.5]$
    
\end{enumerate}

The concept in the first place is the only triadic concept that exactly matches the elements in the query, and for this reason, it received the highest similarity value ($2.67$). The second and third returned concepts also encompass the sought-after elements but with an additional element in the intent ($K$ and $N$ respectively), thus sharing the same value of $2.5$.

When we execute the same query by applying the approach from \citep{Ananias2021}, we get only the concept $(23456,\ R,\ ab)$ as a response, which once again is the top-1 concept returned by our solution. Despite its correct response, we consider that the number of returned concepts by the approximation approach can discourage or even hinder the exploration of concepts similar to a specific user query, requiring the user to make multiple queries to explore a larger set of options.

\subsection{Three-dimensional queries}

The final type of query for matching triadic concepts involves a query where the three components are known. 
The formulation of a three-dimensional query is then $(X_{1}, X_{2}, X_{3})$ and aims at checking if the triple is a triadic concept or identifying the concepts that are the closest to it based on our similarity formula.

\subsubsection{Example}

The following three triadic concepts constitute the top 3 concepts returned for the three-dimensional query $(3,\ R,\ c)$.

\begin{enumerate}
    \item $(16,\ R,\ ac) - [1.17]$
    \item $(23456,\ R,\ ab) - [1.07]$
    \item $(3456,\ KR,\ ab) - [0.92]$
    
\end{enumerate}

As observed, there is no triadic concept that exactly matches the query. However, the most similar concept to this triple is $(16,\ R,\ ac)$, where the sought-after extent is absent, but there is an exact match for the attribute component ($R$) and one additional element $a$ to the requested value of $c$ for the condition component.
The second concept lacks the condition $c$ but exactly possesses the sought-after $R$ attribute and the element $3$ in its object component along with four extra objects, compared to the requested triple.
Lastly, there is the concept $(3456,\ KR,\ ab)$, where the condition $c$ is missing, but its intent contains the extra element $K$, and its extent also includes the three additional objects $4$, $5$, and $6$.

Regarding the result generated by \citep{Ananias2021} for the same query, we obtained the following triadic concepts: $(16,\ R,\ ac)$, $(123456,\ \varnothing,\ abcd)$, and $(23456,\ R,\ ab)$. Comparing the results, we can see that two triadic concepts are present in the output of both approaches: $(16,\ R,\ ac)$ and $(23456,\ R,\ ab)$.

However, the solution from \citep{Ananias2021} also returned $(123456,\ \varnothing,\ abcd)$, a concept that was ranked as the top 5 in our solution. In other words, our method identified two more relevant concepts to present to the user before the concept $(123456,\ \varnothing,\ abcd)$ inside the \textit{supremum} (see Figure \ref{fig:hasse_diagram}), which might not be so interesting to the user.

Furthermore, another crucial aspect to highlight is the significance of ranking the responses, in addition to using a similarity metric. The concepts returned by \citep{Ananias2021} lack any order.
As stated before, the search for a query answer in \citep{Ananias2021} exploits derivation operators and includes sometimes concepts that are in the neighbourhood of the generated concepts. Concept $(123456,\ \varnothing,\ abcd)$ is indeed in the upper cover of $(23456,\ R,\ ab)$ and has the particularity that its extent and its modus cover their counterparts in the second concept.

\section{Empirical study}
\label{Results}

The purpose of this section is to carry out a validation process and provide empirical insights on the execution times of each one of the two solutions: ours and the approximation approach \citep{Ananias2021} using our software platform. The tests were conducted on both real and synthetic data sets.

Given that execution times might exhibit slight variations for the same context across different code runs, the two approaches were executed three times each, and the average execution time was calculated along with the standard deviation.

Furthermore, we conducted a comparative analysis between the algorithms proposed in this study and the approximation solution. The primary goal is to assess the scalability of each approach and evaluate the execution time in large contexts. For all tests, random queries were created in order to compare the solutions.
 

Both algorithms were implemented in Python (3.11), and all empirical tests were carried out on a macOS 13.4.1-based system equipped with 16 GB of RAM and an Apple M2 Pro 10-core processor. 

The experiments are aimed at validating our hypothesis that our solution, based on an inverted index, will exhibit superior scalability and, consequently, improved execution times compared to the approximation approach.

\subsection{Mushroom dataset}

\textit{The Mushroom Data Set}\footnote{Available at: \url{https://archive.ics.uci.edu/ml/datasets/mushroom}} is a well-known one in data mining validation tests. This dataset contains descriptions of hypothetical samples corresponding to 23 species of mushrooms, characterized by 22 variables each one with two to twelve modalities.
In order to have a synthetic triadic context, we first converted the multivariate dataset into a formal context from which we selected 128 binary attributes, decomposed the latter group into a set of 32 attributes and a set of 4 corresponding conditions.
For this experiment, the synthetic triadic context has 8416 objects, 32 attributes and 4 conditions, resulting in a total of 1859 triadic concepts.
Table \ref{tab:res_mushroom_dataset} presents the execution time for the three types of queries, namely one, two, and three-dimension ones, along with the execution time for the data structure creation in both solutions.

\begin{table}[!ht]
\centering
\caption{Mushroom dataset - Execution time comparison}
\label{tab:res_mushroom_dataset}
\resizebox{11cm}{!}{%
\begin{tabular}{l|cc}
                        & \multicolumn{1}{c|}{\textbf{Concept approximation}} & \textbf{Our solution}        \\ \hline
\textbf{Create data structure}   & \multicolumn{1}{c|}{12.2 s ± 481 ms}       & 65.8 ms ± 4.11 ms   \\ \hline
One-dimensional query   & \multicolumn{2}{c}{$(-,-,X_3)$}                                  \\
Execution time          & 3min 9s ± 607 ms                           & 18.5 ms ± 2.7 ms    \\ \hline
One-dimensional query   & \multicolumn{2}{c}{$(-,X_2,-)$}                                  \\
Execution time          & 1.41 s ± 9.47 ms                           & 0.607 ms ± 0,323 ms \\ \hline
One-dimensional query   & \multicolumn{2}{c}{$(X_1,-,-)$}                                  \\
Execution time          & 4.19 ms ± 680 ms                           & 26.9 ms ± 6.66 ms   \\ \hline
Two-dimensional query   & \multicolumn{2}{c}{$(-,X_2,X_3)$}                                \\
Execution time          & 210 ms ± 7.74 ms                           & 2.39 ms ± 0,280 ms  \\ \hline
Two-dimensional query   & \multicolumn{2}{c}{$(X_1,-,X_3)$}                                \\
Execution time          & 2.28 ms ± 0.307 ms                         & 8.11 ms ± 0.815 ms  \\ \hline
Two-dimensional query   & \multicolumn{2}{c}{$(X_1,X_2,-)$}                                \\
Execution time          & 48 ms ± 3.44 ms                            & 5.64 ms ± 0.922 ms  \\ \hline
Three-dimensional query & \multicolumn{2}{c}{$(X_1,X_2,X_3)$}                              \\
Execution time          & 257 ms ± 9.95 ms                           & 28.8 ms ± 4.49 ms  
\end{tabular}%
}
\end{table}

The first noticeable point is the execution time for creating the data structure used to perform the queries. In the approximation approach, the transformation of the triadic context into three dyadic contexts is required, which involves calculating the Cartesian product of two dimensions out of three in the original context. Thus, just the Cartesian product between the set of objects and the set of attributes resulted in a context with 269,312 cells in the dyadic context $\KK^{(3)}$. Consequently, this demands a longer execution time, averaging 12.2 seconds for creating this data structure.

In contrast, for creating the inverted index, it is only necessary to traverse the list of triadic concepts once to get the concepts associated with each dimension element. In this way, we have an average execution time of 65.8 milliseconds for its creation.

Regarding the execution of queries, the run time for both algorithms was measured in milliseconds, with the exception of the first two queries in the solution of \citep{Ananias2021}. For the first one-dimensional query, all derivation operations are performed on the context $\KK^{(3)}$ with 269,312 cells. Furthermore, the derivation of a single element can generate a large set of pairs, which subsequently undergoes a factorization process. This excessive amount of processing results in an average execution time of 3 minutes and 9 seconds, while our solution runs in an average of 18.5 milliseconds.

In the second one-dimensional query, the number of cells in the generated dyadic context is also a bottleneck. For this query, the derivations are performed on the context $\KK^{(2)}$ with 33,664 cells, resulting in an average time of 1.41 seconds. On the other hand, our solution took only 0.607 milliseconds on average.

It is also important to highlight the number of triadic concepts returned by each solution. In all queries, our approach was able to identify a higher number of concepts with their similarity score. In the approximation approach, the size of the query answer depends on the query and the TC set without any hint from the user. Moreover, the returned concepts are not ranked, requiring manual and meticulous exploration from the user to identify the most relevant ones.

\subsection{Groceries dataset}

The \textit{Groceries} database\footnote{Available at: \url{https://www.kaggle.com/heeraldedhia/groceries-dataset}} contains 38765 transactions, 3898 customers, 167 products (items) and 728 distinct transaction dates.
The dataset we are analyzing contains transactions made by customers who bought a set of items during a given month (rather than a specific date) between 2014 and 2015 and has a total of 17456 triadic concepts.
Table \ref{tab:res_groceries_dataset} displays the execution times for each potential query type, namely one-dimensional, two-dimensional, and three-dimensional queries, as well as the time taken for data structure creation in both approaches.

\begin{table}[!ht]
\centering
\caption{Groceries dataset - Execution time comparison}
\label{tab:res_groceries_dataset}
\resizebox{11cm}{!}{%
\begin{tabular}{l|cc}
                        & \multicolumn{1}{c|}{\textbf{Concept approximation}} & \textbf{Our solution}       \\ \hline
\textbf{Create data structure}   & \multicolumn{1}{c|}{1 min 27s ± 456 ms}     & 21.9 ms ± 3.09 ms  \\ \hline
One-dimensional query   & \multicolumn{2}{c}{$(-,-,X_3)$}                                 \\
Execution time          & 58.2 s ± 232 ms                            & 23.9 ms ± 3.94 ms  \\ \hline
One-dimensional query   & \multicolumn{2}{c}{$(-,X_2,-)$}                                 \\
Execution time          & 429 ms ± 18.6 ms                           & 4.53 ms ± 0.313 ms \\ \hline
One-dimensional query   & \multicolumn{2}{c}{$(X_1,-,-)$}                                 \\
Execution time          & 3.87 ms ± 0.873 ms                         & 2.05 ms ± 0.375 ms \\ \hline
Two-dimensional query   & \multicolumn{2}{c}{$(-,X_2,X_3)$}                               \\
Execution time          & 48.8 ms ± 7.02 ms                          & 57.5 ms ± 8.56 ms  \\ \hline
Two-dimensional query   & \multicolumn{2}{c}{$(X_1,-,X_3)$}                               \\
Execution time          & 1.65 s ± 5.02 ms                           & 33.6 ms ± 5.59 ms  \\ \hline
Two-dimensional query   & \multicolumn{2}{c}{$(X_1,X_2,-)$}                               \\
Execution time          & 40.8 ms ± 8 ms                             & 6.02 ms ± 0.779 ms \\ \hline
Three-dimensional query & \multicolumn{2}{c}{$(X_1,X_2,X_3)$}                             \\
Execution time          & 6.74 s ± 9.88 ms                           & 92.9 ms ± 2.59 ms 
\end{tabular}%
}
\end{table}

In this experiment, we can observe that the approximation approach, once again, requires more time for data structure creation. In this case, the fact that we have a larger number of objects, attributes, and conditions has a negative impact on the creation of dyadic contexts, requiring an average time of one minute and 27 seconds for its completion. In contrast, the creation of the inverted index is not significantly affected, even with a considerably large number of triadic concepts in this dataset, taking only 21.9 milliseconds to create the index.

Using this dataset, the limitation of the approximation approach becomes evident, where three queries took more than 1.5 seconds to return a response, with the slowest one being the first query in Table \ref{tab:res_groceries_dataset}, which took an average of 58.2 seconds. Once again, the number of cells in the dyadic context $\KK^{(2)}$ generated by the Cartesian product of objects and attributes (650,966 cells), along with the derivation and factorization operations, prove to be non scalable and impractical for larger contexts.


This dataset led to a large set of 17,456 triadic concepts. However, our solution was efficient to find the concepts using the inverted index as discussed earlier.

\section{Conclusion}
\label{Conclusion}

In this paper, we introduce a novel approach to querying triadic concepts based on partial or complete matching of triples. This approach, based on using an inverted index, does not need to store or explore any context, nor rely on derivation or factorization operations to identify the triadic concepts most similar to a given query.

This solution not only shows higher efficiency compared to the most related existing approach in the literature but also indicates enhanced scalability, making it applicable in big data scenarios. To the best of our knowledge, there are no other algorithms or methods capable of partial or complete matching of triples without the need to perform derivation operators or to transform the triadic context into dyadic ones. Moreover, the query answer sounds close to the user's triple.

Currently, we are conducting additional tests and code improvement for the two compared approaches as well as working on an open-source platform for exploring and mining patterns in Triadic Concept Analysis.

\bibliographystyle{unsrtnat}
\bibliography{Refs}

\begin{thebibliography}{11}
\providecommand{\natexlab}[1]{#1}
\providecommand{\url}[1]{\texttt{#1}}
\expandafter\ifx\csname urlstyle\endcsname\relax
  \providecommand{\doi}[1]{doi: #1}\else
  \providecommand{\doi}{doi: \begingroup \urlstyle{rm}\Url}\fi

\bibitem[Kwuida et~al.(2010)Kwuida, Missaoui, Amor, Boumedjout, and Vaillancourt]{Kwuida2010}
L{\'{e}}onard Kwuida, Rokia Missaoui, Beligh~Ben Amor, Lahcen Boumedjout, and Jean Vaillancourt.
\newblock Restrictions on concept lattices for pattern management.
\newblock In Marzena Kryszkiewicz and Sergei~A. Obiedkov, editors, \emph{Proceedings of the 7th International Conference on Concept Lattices and Their Applications, Sevilla, Spain, October 19-21, 2010}, volume 672 of \emph{{CEUR} Workshop Proceedings}, pages 235--246. CEUR-WS.org, 2010.

\bibitem[Ganter and Wille(1999)]{Ganter1999}
Bernhard Ganter and Rudolf Wille.
\newblock \emph{Formal Concept Analysis: Mathematical Foundations}.
\newblock Springer Berlin Heidelberg, Berlin, Heidelberg, jun 1999.
\newblock ISBN 978-3-540-62771-5.

\bibitem[Lehmann and Wille(1995)]{Lehmann1995}
Fritz Lehmann and Rudolf Wille.
\newblock {A triadic approach to formal concept analysis}.
\newblock In \emph{Lecture Notes in Computer Science}, volume 954, pages 32--43, 1995.
\newblock ISBN 3540601619.

\bibitem[J{\"a}schke et~al.(2008)J{\"a}schke, Hotho, Schmitz, Ganter, and Stumme]{Jaschke2008}
Robert J{\"a}schke, Andreas Hotho, Christoph Schmitz, Bernhard Ganter, and Gerd Stumme.
\newblock {Discovering shared conceptualizations in folksonomies}.
\newblock \emph{Web Semantics}, 6\penalty0 (1):\penalty0 38--53, 2008.
\newblock ISSN 15708268.

\bibitem[Missaoui et~al.(2020)Missaoui, Ruas, Kwuida, and Song]{Missaoui2020}
Rokia Missaoui, Pedro H.~B. Ruas, L{\'e}onard Kwuida, and Mark A.~J. Song.
\newblock Pattern discovery in triadic contexts.
\newblock In Mehwish Alam, Tanya Braun, and Bruno Yun, editors, \emph{Ontologies and Concepts in Mind and Machine}, pages 117--131, Cham, 2020. Springer International Publishing.
\newblock ISBN 978-3-030-57855-8.

\bibitem[Rudolph et~al.(2015)Rudolph, S{\u{a}}c{\u{a}}rea, and Troanc{\u{a}}]{Rudolph2015}
Sebastian Rudolph, Christian S{\u{a}}c{\u{a}}rea, and Diana Troanc{\u{a}}.
\newblock Towards a navigation paradigm for triadic concepts.
\newblock In Jaume Baixeries, Christian Sacarea, and Manuel Ojeda-Aciego, editors, \emph{Formal Concept Analysis}, pages 252--267, Cham, 2015. Springer International Publishing.
\newblock ISBN 978-3-319-19545-2.

\bibitem[Baixeries et~al.(2009)Baixeries, Szathmary, Valtchev, and Godin]{Baixeries2009}
Jaume Baixeries, Laszlo Szathmary, Petko Valtchev, and Robert Godin.
\newblock Yet a faster algorithm for building the hasse diagram of a concept lattice.
\newblock In S{\'e}bastien Ferr{\'e} and Sebastian Rudolph, editors, \emph{Formal Concept Analysis}, pages 162--177, Berlin, Heidelberg, 2009. Springer Berlin Heidelberg.
\newblock ISBN 978-3-642-01815-2.

\bibitem[Wille(1995)]{Wille95}
Rudolf Wille.
\newblock The basic theorem of triadic concept analysis.
\newblock \emph{Order}, 12\penalty0 (2):\penalty0 149--158, 1995.

\bibitem[Zobel and Moffat(2006)]{Zobel2006}
Justin Zobel and Alistair Moffat.
\newblock Inverted files for text search engines.
\newblock \emph{ACM Comput. Surv.}, 38\penalty0 (2):\penalty0 6–es, jul 2006.
\newblock ISSN 0360-0300.

\bibitem[Ananias et~al.(2021)Ananias, Missaoui, Ruas, Zarate, and Song]{Ananias2021}
Kaio~H.A. Ananias, Rokia Missaoui, Pedro~H.B. Ruas, Luis~E. Zarate, and Mark~A.J. Song.
\newblock Triadic concept approximation.
\newblock \emph{Information Sciences}, 572:\penalty0 126--146, 2021.

\bibitem[Kis et~al.(2016)Kis, Sacarea, and Troanca]{Kis2016}
Levente~Lorand Kis, Christian Sacarea, and Diana Troanca.
\newblock Fca tools bundle-a tool that enables dyadic and triadic conceptual navigation.
\newblock In \emph{FCA4AI@ ECAI}, pages 42--50, 2016.

\end{thebibliography}

\end{document}